\documentclass{aastex63}
\usepackage{color,soul}

\usepackage{graphicx}%,caption,subcaption}

\submitjournal{ApJ}

\shortauthors{Malik, Banik \& Bandyopadhyay}

\graphicspath{{./}{figures/}}

\begin{document}

\title{Equation of State table with hyperon and antikaon  
for supernova and neutron star merger%\footnote{Released on June, 10th, 2019}
}

\correspondingauthor{Sarmistha Banik}
\email{sarmistha.banik@hyderabad.bits-pilani.ac.in}

\author[0000-0003-2633-5821]{Tuhin Malik}
\affiliation{Birla Institute of Technology \& Sciences, Pilani\\
Department of Physics, Hyderabad-500078, India}

\author[0000-0003-0221-3651]{Sarmistha Banik}
\affiliation{Birla Institute of Technology \& Sciences, Pilani\\
Department of Physics, Hyderabad-500078, India}

\author[0000-0003-0616-4367]{Debades Bandyopadhyay}
\affiliation{Saha Institute of Nuclear Physics, HBNI, 
1/AF Bidhannagar, Kolkata-700064, India}
\affil {Frankfurt Institute for Advanced Studies (FIAS)\\
Ruth Moufang Strasse 1,
D-60438 Frankfurt am Main, Germany}

\begin{abstract}

We develop a new equation of state (EoS) table involving thermal (anti)kaons, Bose-Einstein condensate of $K^{-}$ mesons and $\Lambda$-hyperons for core-collapse supernova  and neutron star merger simulations. This EoS table is based on a finite temperature density-dependent relativistic hadron field theory where baryon-baryon interaction is mediated by scalar $\sigma$, vector $\omega$ and
$\rho$ mesons, using the parameter set DD2 for nucleons. The repulsive hyperon-hyperon interaction is mediated by an additional  strange  $\phi$ meson. The EoS for the $K^-$ condensed matter is also calculated within the framework of relativistic mean field model, whereas the low-density, inhomogeneous matter is calculated in the extended Nuclear Statistical Equilibrium model (NSE). The EoS
table is generated for a wide range of values of three parameters - baryon
density ($10^{-12}$ to $\sim$
1 fm$^{-3}$), positive charge fraction(0.01 to 0.60) and temperature(0.1 to 158.48 MeV).

\end{abstract}


%% We recommend that authors also use the natbib \cite

\section{Introduction} %\label{sec:intro}
The study of the equation of state of matter in neutron stars has reached a high level of sophistication observationally  as well as theoretically in recent years. This journey had started with the discovery of the first pulsar \cite{bell} in the year 1967. In recent years, it has got huge boost with the detection of gravitational waves in binary neutron star (BNS) merger event GW170817 \cite{ligo1,ligo2,ligo3,ligo4,ligo5,ligo6}. Parallelly, the observations of heavy neutron stars led to the determination of masses 
$\geq 2 M_{solar}$ from post Keplerian parameters using the pulsar timing technique \cite{anto,croma}. The 
simultaneous measurement of mass and radius of the x-ray 
powered pulsar PSR J0030+0451 is another important step forward by the Neutron Star Composition Explorer (NICER) \cite{riley,miller,watts}. All these observations are providing valuable inputs to constrain the EoS from low to very high density which could not be otherwise probed using the knowledge of laboratory experiments on nuclei and relativistic heavy ion collisions.

Observations of galactic massive pulsars PSR J0348+0432 of 2.01$\pm 0.04$ M$_{solar}$, PSR J0740+6620 of 2.14$^{+0.10}_{-0.09}$ M$_{solar}$ and PSR J1810+1744 of
$2.13 \pm 0.04$ M$_{solar}$ set a lower limit on the maximum mass of neutron stars \cite{anto,croma,romani}. On the other hand, the matter ejected in the BNS merger GW170817 was also observed across the wide electromagnetic spectrum. An upper bound on the maximum mass
of neutron stars might be obtained from the electromagnetic observation of GW170817 if the remnant collapsed to a black hole. Different groups estimated the upper bound to be in the range $\sim 2.17 - 2.3 M_{solar}$ \cite{metz,luci,shapiro,shibata,enping}. Furthermore, it was possible for the first time to extract the value of tidal deformability (${\widetilde{\Lambda}}$) from the gravitational waves signal of GW170817.  Its range of $70\leq \widetilde{\Lambda} \leq 720$ gives an estimate for the radius as 9-14 km for a 1.4 M$_{solar}$ neutron star 
\cite{ligo3,sch18,fattoyev,ozel,de,zhao,soma}. The values of mass-radius of PSR J0030+0451 from two different analyses of NICER results are  1.44$^{+0.15}_{-0.14}$ M$_{solar}$-13.02$^{+1.24}_{-1.06}$ km  and 1.34$^{+0.15}_{-0.16}$M$_{solar}$
%-radius 
-12.71$^{+1.14}_{-1.19}$ km \cite{miller,riley}. Another compact binary coalescence event
GW190814 involving a black hole of $23.2^{+1.1}_{1.0} M_{solar}$ and a compact object of 
$2.59^{+0.8}_{0.9} M_{solar}$ had been reported recent \cite{ligo7}. This event has 
generated a debate whether the secondary mass makes the compact object the heaviest neutron 
star or a lightest black hole.  
%All these i
The observed values of masses and radii are stringent probes of the EoS of neutron star matter. Theoretical models of EoS should be compatible with these
observations. 

The theoretical modeling of EoS has undergone a sea of changes over the past several years. 
Traditional EoS models are based on two-body plus three-body interactions in non-relativistic approaches and the strong interaction Lagrangian in relativistic field theoretic approaches \cite{ls,Shen}. Furthermore, compositions of matter are assumed in both types of models. Recently, large numbers of EoSs were constructed by adopting phenomenological Skyrme interaction or nucleon-nucleon chiral potentials for the low density; and perturbative Quantum Chromodynamics for asymptotically high density regimes. The EoS in the intermediate density regime is connected to EoSs at two extreme densities either by polytropes or just demanding the causality and monotonicity \cite{heb,kurk,sch18,alford}. However, such EoS models do not assume any kind of composition of matter in the intermediate density regime which plays the most important role for masses and radii of neutron stars. 

Traditional models of EoS are widely used in
numerical relativity simulations of compact astrophysical objects. Our motivation in this work is to compute
an EoS including strange matter within the framework of relativistic field theoretical models to be used an input in
core collapse supernova (CCSN) and neutron star merger simulations.   

A large set of EoSs with and without strange matter such as hyperon, antikaon condensed or quark matter are already available for CCSN and neutron star merger simulations 
\cite{hs1,raduta10,horo,horo1,horo2,fis1,bli,hs2,stei,buyu,composemanual,fis2,toga,const,bhb,rmp,EPJC}. The EoS tables with nucleons-only matter satisfy the 2 M$_{solar}$ lower bound on the maximum mass of neutron stars. However, many EoSs with hyperons or quarks used
in CCSN simulations do not conform to the lower bound on the maximum mass \cite{ishi,naka08,irina,sumi,shen11,naka,oertel12,peres,sb}. Keeping this in mind, we constructed an EoS table including $\Lambda$ hyperons  which is compatible with 
2 M$_{solar}$ neutron stars\cite{bhb}. This EoS, known as BHB$\Lambda\phi$, is being widely used in CCSN \cite{Char2015} and BNS merger simulations \cite{Radicebns}. It has been long debated that the Bose-Einstein condensate of negatively charged kaons might appear in dense matter \cite{kap,knor}. This idea was extended to understand the non-observation of a neutron star in SN1987A \cite{bethe}. There was no EoS table involving an antikaon condensate in nuclear 
matter before our very recent work on this problem \cite{EPJC}. We did not 
consider the 
appearance of hyperons along with the antikaon condensate in the previous work.
%If hyperons appears first, it would
The early appearance of hyperons would delay the onset of Bose-Einstein condensate of antikaons in dense matter or vice versa. It 
is worth investigating how one form of strange matter behaves in presence of another.
%another form of strange matter. 
This motivates us to investigate this issue and extend our hyperon 
EoS BHB$\Lambda\phi$ to include the Bose-Einstein condensate of $K^-$ mesons fulfilling most 
updated information on neutron stars.

The paper is organised as follows. In Section 2, the hadronic field theory models of EoSs at zero and finite temperatures are described. The results of our calculation are discussed in Section 3. Section 4 contains the summary and conclusions.

\section{The Model}
\subsection{Hadronic Model}
The baryonic matter is described within the framework of the density-dependent model
adopting the relativistic mean field (RMF) approximation.
The baryon-baryon interaction is mediated by $\sigma$, $\omega$,
$\rho$ and $\phi$ mesons. However,  nucleons
do not couple to $\phi$ mesons, they account for the repulsive hyperon-hyperon interaction. The Lagrangian density is given by \cite{typ10,bhb}:

\begin{eqnarray}\label{had}
{\cal L} &=& \sum_{B=N,\Lambda} \bar\Psi_{B}\left(i\gamma_\mu{\partial^\mu} - m_B
+ g_{\sigma B} \sigma - g_{\omega B} \gamma_\mu \omega^\mu
- g_{\rho B}
\gamma_\mu{\mbox{\boldmath $\tau$}}_B \cdot
{\mbox{\boldmath $\rho$}}^\mu
- g_{\phi B} \gamma_\mu \phi^\mu
\right)\Psi_B\nonumber\\
&& + \frac{1}{2}\left( \partial_\mu \sigma\partial^\mu \sigma
- m_\sigma^2 \sigma^2\right)
 -\frac{1}{4} \omega_{\mu\nu}\omega^{\mu\nu}
+\frac{1}{2}m_\omega^2 \omega_\mu \omega^\mu 
 - \frac{1}{4}{\mbox {\boldmath $\rho$}}_{\mu\nu} \cdot
{\mbox {\boldmath $\rho$}}^{\mu\nu}
+ \frac{1}{2}m_\rho^2 {\mbox {\boldmath $\rho$}}_\mu \cdot
{\mbox {\boldmath $\rho$}}^\mu
 -\frac{1}{4} \phi_{\mu\nu}\phi^{\mu\nu}
+\frac{1}{2}m_\phi^2 \phi_\mu \phi^\mu~,
\label{lagm}
\end{eqnarray}
where $\Psi_B$ denotes the isospin multiplets for baryons B,  $m_B$'s being their bare masses, and ${\tau _{B}}$, the isospin operator. The density dependent
meson-nucleon couplings are denoted by $g_{x B}$ with x = $\sigma$, $\omega$, and  $\rho$  
meson fields. The vector meson field strength tensors are particularly represented by   $x^{\mu \nu} = \partial^ \mu x^\nu -\partial^ \nu x^\mu$. In the mean field approximations, meson fields are replaced by their expectation values $\left<x\right>$. Only the time-like components of vector fields and the third isospin component of the $\rho$ field survive in a uniform and static matter. They are denoted by $\sigma$, $\omega_0$, $\rho_{03}$ and $\phi_0$ and are obtained by solving the meson field equations in the RMF approximation,
$$m_{\sigma}^2\sigma = \sum_{B}g_{\sigma B}
\bar\psi_{\sigma B}\psi_{B}, \quad 
m_{\omega}^2\omega_0 = \sum_{B}g_{\omega B}
\bar\psi_{B}\gamma_{0}\psi_{B}, \quad 
{m_\rho}^{2}\rho_{03}={1\over2}\sum_{B}
g_{\rho B}\bar\psi_{B}\gamma_{0}\tau_{3B} \psi_{B}, \quad m_{\phi}^2\phi_0 = \sum_{B}g_{\phi B}
\bar\psi_{B}\gamma_{0}\psi_{B}.$$  
The grand-canonical thermodynamic potential per unit volume of the hadronic phase is given by \cite{bhb,soma}
\begin{eqnarray}
\frac{\Omega}{V} &=& \frac{1}{2}m_\sigma^2 \sigma^2
- \frac{1}{2} m_\omega^2 \omega_0^2 
- \frac{1}{2} m_\rho^2 \rho_{03}^2  
- \frac{1}{2} m_\phi^2 \phi_0^2 
- \Sigma^r \sum_{B=n,p,\Lambda} n_B
\nonumber \\
&& - 2T \sum_{B=n,p,\Lambda} \int \frac{d^3 k}{(2\pi)^3} 
[\mathrm{ln}(1 + e^{-\beta(E^* - \nu_B)}) +
\mathrm{ln}(1 + e^{-\beta(E^* + \nu_B)})] ~,  
\label{gctpot}
\end{eqnarray}
where the temperature is defined as $\beta = 1/T$ and
$E^* = \sqrt{(k^2 + m_B^{*2})}$.

The number densities and scalar number densities of baryon B are   $n_B= <\bar\psi_{B}\gamma_{0} \psi_{B}>$,  and $n_B^S=<\bar\psi_{\sigma B}\psi_{B}>$ respectively, and at finite temperature (T) are given by,
 \begin{equation}
n_B = 2 \int \frac{d^3k}{(2\pi)^3} \left({\frac{1}{e^{\beta(E^*-\nu_B)} 
+ 1}} - {\frac{1}{e^{\beta(E^*+\nu_B)} + 1}}\right), 
\end{equation}

\begin{equation}
n_B^S = 
2 \int \frac{d^3 k}{(2\pi)^3} \frac{m_B^*}{E^*} 
\left({\frac{1}{e^{\beta(E^*-\nu_B)} 
+ 1}} + {\frac{1}{e^{\beta(E^*+\nu_B)} + 1}}\right) ~.
\end{equation}
Here the effective nucleon mass is
$m^{*}_{B} = m_B - g_{\sigma B} \sigma$.
The chemical potential is given by
$\mu_{B} = \nu_B + g_{\omega B} \omega_0 + g_{\rho B} \tau_{3B} \rho_{03}+ g_{\phi B} \phi_0  + \Sigma^r~,$
where the rearrangement term $\Sigma^r$ takes care of many-body effects in nuclear interaction \cite{typ10,bhb}.  It arises due to the density-dependence of the couplings and is expressed as
\begin{equation}
\Sigma^r = 
\sum_B[-\frac {\partial g_{\sigma B}} {\partial n_B}  
\sigma n^{s}_B + \frac {\partial g_{\omega B}} {\partial n_B}  \omega_0 n_B
+ \frac {\partial g_{\rho B}} {\partial n_B}\tau_{3N} \rho_{03} n_B + \frac {\partial g_{\phi B}} {\partial n_B}  \phi_0 n_B]~.
\end{equation}
The pressure  is calculated from the grand-canonical thermodynamic potential per unit volume as $P = - {\Omega}/V$ as given by Eq.(\ref{gctpot}) and includes the rearrangement term.

However, the  energy density does not explicitly contain the 
rearrangement term, as evident from its expression,
\begin{equation}\label{eqn:ed}
\epsilon_B = \frac{1}{2}m_\sigma^2 \sigma^2
+ \frac{1}{2} m_\omega^2 \omega_0^2
+\frac{1}{2} m_\rho^2 \rho_{03}^2 + \frac{1}{2} m_\phi^2 \phi_0^2
 + 2 \sum_{B=N, \Lambda} \int \frac{d^3 k}{(2\pi)^3} E^* 
\left({\frac{1}{e^{\beta(E^*-\nu_B)} 
+ 1}} + {\frac{1}{e^{\beta(E^*+\nu_B)} + 1}}\right)~.  
\end{equation}
The rearrangement term  not only accounts for the energy-momentum conservation but also the  thermodynamic consistency of the system. The energy density and pressure are related through Gibbs-Duhem relation i.e. ${\mathcal S}_B = \beta \left(\epsilon_B + P_B - \sum_{B} \mu_B n_B \right)$, where,  ${\mathcal S}_B$ is the entropy density of baryon B. 
\subsection{Model for antikaons}
Kaon-nucleon interaction is considered in the same footing as the
nucleon-nucleon interaction in Eq.(\ref{lagm}). The Lagrangian density for
(anti)kaons in the minimal coupling scheme is \cite{glen99,pons,banik01,sb08,char14},
\begin{equation}
{\cal L}_K = D^*_\mu{\bar K} D^\mu K - m_K^{* 2} {\bar K} K ~,
\end{equation}
where $K$ and $\bar K$ denote kaon and (anti)kaon doublets; the covariant
derivative is
$D_\mu = \partial_\mu + ig_{\omega K}{\omega_\mu}
+ i g_{\rho K}
 \tau_K \cdot  \rho_\mu +ig_{\phi K}{\phi_\mu}$ and
the effective mass of antikaon is $m_K^* = m_K - g_{\sigma K} \sigma$.
Meson fields are modified in the presence of $K^-$ condensate \cite{glen99,banik01}
%. The rearrangement term  implicitly changes via  meson fields which 
and are obtained solving the following equations:
 $$m_{\sigma}^2\sigma = \sum_{B}g_{\sigma B}
n_B^S +\sum_{K}g_{\sigma K} n_K,  \quad
m_{\omega}^2\omega_0 = \sum_{B}g_{\omega B}
n_B -\sum_{K}g_{\sigma K} n_B,$$
$${m_\rho}^{2}\rho_{03}={1\over2}\sum_{B}
g_{\rho B}\tau_{3B}n_B +\sum_{K}g_{\sigma K} n_K\tau_{3K}, \quad  m_{\phi}^2\phi_0 = \sum_{B}g_{\phi B}
n_B-\sum_{K}g_{\sigma K}
n_K.$$  
The thermodynamic potential for (anti)kaons is given by \cite{pons,sb08},
\begin{equation}
\frac {\Omega_K}{V} = T \int \frac{d^3k}{(2\pi)^3} [ ln(1 -
e^{-\beta(\omega_{K^-} - \mu)}) +
 ln(1 - e^{-\beta(\omega_{K^+} + \mu)})]~,
\end{equation}
where the in-medium energies of $K^{\pm}$ mesons are given by
\begin{equation}
\omega_{K^{\pm}} =  \sqrt {(k^2 + m_K^{*2})} \pm \left( g_{\omega K} \omega_0
+ \frac {1}{2} g_{\rho K} \rho_{03} +g_{\phi K} \phi_0 \right)~.
\end{equation}
 The chemical potentials of nucleons are related to that of $K^-$ mesons  by $\mu = \mu_n -\mu_p$ \cite{pons}. For s-wave (${\bf k}=0$) condensation, the momentum dependence vanishes in $\omega_{K^\pm}$. The in-medium energy of $K^-$ condensate decreases from its vacuum value $m_K$ as the meson fields build up with increasing density.  The $K^-$ condensate appears in the system as $\omega_{K^-}$ equals to its chemical potential i.e. 
$\mu = \omega_{K^{-}} =   m_K^* - g_{\omega K} \omega_0 
- \frac {1}{2} g_{\rho K} \rho_{03}- g_{\phi K} \phi_0$~.  This is the threshold condition for $K^-$ condensation.   Incidentally, the threshold condition for $K^+$ condition $\mu =\omega_{K^+}$ is never attained because of the repulsive nature of  $K^+$-nucleon interaction. We calculate the pressure due to thermal (anti)kaons using $P_K = -{\Omega_K}/{V}$. $K^-$ condensate does not contribute to pressure. The energy density of (anti)kaons is due to the condensate as well as the thermal (anti)kaons and is  given by, 
\begin{equation}\label{eq_epsilonK}
\epsilon_K = m_K^* n_K^C + \left( g_{\omega K} \omega_0
+ \frac {1}{2} g_{\rho K} \rho_{03} +g_{\phi K} \phi_0 \right) n_K^T \nonumber +
\int \frac{d^3 k}{(2\pi)^3} 
\left({\frac{\omega_{K^-}}{e^{\beta(\omega_{K^-}-\mu)} 
- 1}} + {\frac{\omega_{K^+}}{e^{\beta(\omega_{K^+}+\mu)} - 1}}\right)~.  
\end{equation}
The thermodynamic quantities-entropy density, energy density, pressure, chemical potential and number density are related to each other through  
${\mathcal S}_K = \beta \left(\epsilon_K + P_K -  \mu n_K \right)$. The entropy per baryon is given by $s={\mathcal S}/n_B$, where $n_B$ is the total baryon density.
The total (anti)kaon number density ($n_K$)  is given by, $n_K = n_K^C + n_K^{T}~$,
where, $n_K^C$ and and $n_K^{T}$ are the $K^-$ condensate  and  thermal (anti)kaon
density respectively. They are  given by, 
\begin{eqnarray}
n^C_{K} &=& 2\left( \omega_{K^-} + g_{\omega K} \omega_0
+ \frac{1}{2} g_{\rho K} \rho_{03} +g_{\phi K} \phi_0 \right) {\bar K} K
= 2m^*_K {\bar K} K  ~, \nonumber \\
n_K^{T}& = &
\int \frac{d^3k}{(2\pi)^3}
\left({\frac{1}{e^{\beta(\omega_{K^-}-\mu)}
- 1}} - {\frac{1}{e^{\beta(
\omega_{K^+}+\mu)} - 1}}\right)~.
\end{eqnarray}

\subsection{Parameters of the DD2 model}

The nucleon-meson couplings depend on the density through $g_{\alpha B}(n_B)=g_{\alpha B}(n_{0}) f_{\alpha}(x)$ where  $x=n_B/n_0$,  and 
$$f_{\alpha}(x)=a_{\alpha}\frac{1+b_{\alpha}(x+d_{\alpha})^2}{1+c_{\alpha}
(x+d_{\alpha})^2}$$ for $\alpha=\omega$, $\sigma$ \cite{typ99,typ10}. 
An exponential density-dependence is assumed  for the isovector meson ${\rho}$ i.e. 
$f_{\alpha}(x) = \exp[{-a_{\alpha} (x-1)}]$, as their couplings decrease at higher densities
 \cite{typ99}.
The DD2 parameter set of nucleon-meson couplings
used in the calculation, results in the saturation properties of symmetric
nuclear matter such as $n_0 =$ 0.149065 fm$^{-3}$, binding energy per nucleon
16.02 MeV,
incompressibility 242.7 MeV, symmetry energy 31.67 MeV and its slope coefficient
55.03 MeV \cite{typ10,fis2}. 

The system is populated with $\Lambda$'s when the chemical equilibrium
condition $\mu_n = \mu_\Lambda$ is met. The density-dependent $\Lambda$-vector meson hyperon vertices are obtained
from the SU(6)
symmetry of the quark model\cite{dov,sch96}. $$g_{\omega \Lambda} = \frac{2}{3} g_{\omega N}, \quad g_{\rho \Lambda} = 0, \quad g_{\phi \Lambda} = -\frac{\sqrt 2}{3} g_{\omega N}.$$  On the other hand, the $\Lambda$ hyperon - scalar meson couplings are obtained from the hypernuclei data.  We consider $\Lambda$ hyperon potential depth  $U_{\Lambda}^N = g_{\omega \Lambda}\omega_0 - g_{\sigma \Lambda} \sigma_0 + {\Sigma}^{r}= -30$ MeV in normal nuclear
matter \cite{mil,sch92,mar} and the ratio of $g_{\sigma \Lambda}$ to $g_{\sigma N}$ is 0.62008.

However, kaon-meson
couplings are not density-dependent. The kaon-vector meson coupling 
constants are also estimated exploiting the quark model and isospin counting rule
i.e. $g_{\omega K} = \frac{1}{3} g_{\omega N}$ and $g_{\rho K} = g_{\rho N}$
\cite{sch96,banik01}. The scalar coupling constant is determined from the real
part of $K^{-}$ optical potential $U_{K^{-}} = - g_{\sigma K}\sigma_0 - g_{\omega K} \omega_0 + {\Sigma}^{r}~$ at the saturation density, for which  an appropriate value of -120 MeV is considered in this work such that the cold beta-equilibrated EoS is compatible with 2 M$_{solar}$ neutron stars. The $K^-$ optical potential could actually range from $-60$ MeV to $-200$ MeV as indicated by the
unitary chiral model calculations and phenomenological fit to kaonic atom data
\cite{Fri94,Fri99,Tol,Tol2}. 

\subsection{Matching with the low density EoS}
The inhomogeneous nuclear matter at low temperatures
($\sim 10 $ MeV) and the sub-saturation density, is composed of  light and
heavy nuclei along with unbound nucleons. In this case, we use the HS(DD2) EoS described within the extended Nuclear Statistical Equilibrium(NSE) Model of Hempel \& Schaffner-Bielich \cite{hs1}.  It is expected that hyperons and/or (anti)kaons would not appear at the low densities and low temperatures. In this situation, we compare the free energy per baryon at fixed $n_B$, T and $Y_q$ of the data of our uniform EoS table with those of the HS(DD2) EoS table. If the free energy of HS(DD2) table is less than that our EoS table, we replace the corresponding data of the latter with those of the former to get the final EoS table with
$\Lambda$ hyperons and (anti)kaons. Furthermore, we impose the condition that hyperons and (anti)kaons are considered only when their fractions are $> 10^{-5}$ to avoid unphysical situations and ensure a smooth transition between low and high densities.  Henceforth
we call this merged EoS table
\footnote{https://universe.bits-pilani.ac.in/Hyderabad/sbanik/EoS}
with $\Lambda$-hyperon and $K^{-}$ as BHB$\Lambda K^- \phi$  EoS. Incidentally we generated a similar EoS table (BHB$\Lambda\phi$) with 
$\Lambda$-hyperons following the similar procedure \cite{bhb}.

\subsection{Accuracy and Consistency of the EoS table}

The following consistency checks on the EoS table are performed.
Thermodynamic consistency is achieved by the condition: 
$$f = \mu_n n_n + \mu_p n_p + \mu_{\Lambda} n_{\Lambda} + \mu n_K - P,$$
where $f = \epsilon - T {\mathcal S}$.

The modulus of the relative thermodynamic accuracy in this EoS table is given by

$$\Delta = \frac {T {\mathcal S} - P + \mu_n n_n + \mu_p n_p+ \mu_{\Lambda} n_{\Lambda} 
+ \mu n_K} {\epsilon} - 1 \sim 10^{-7}.$$

Sum rule of particle fractions ($X_i$) is satisfied by the EoS table given by
$$X_n + X_p + X_s + X_A + X_{\Lambda} = 1.$$

Finally, the EoS table fulfills the thermodynamic stability criteria:
$$\frac{d{\mathcal S}}{dT} > 0,\quad \frac{dP}{dn_B} > 0.~$$

\section{Results and discussion}
The EoS table with  thermal (anti)kaons and $K^{-}$ condensate are generated using the DD2 parameter set and $U_{K^{-}}= - 120$ MeV for baryon densities ($n_B$=$10^{-12}$ to $\sim 1 fm^{-3}$
),
temperatures (T=0.1 to 158.48 MeV) and positive charge
fractions $Y_q(n_B) = n_p - n_K$  ($Y_q$=0.01 to 0.60).
Grid spacing for baryon density is $\Delta log10(n_B) = 0.04$;
for temperature $\Delta log10(T) = 0.04$ and for positive charge
fraction $\Delta Y_q =0.01$. The EoS table consists of  301 baryon density points,
81 temperatures and 60 positive charge fractions i.e. total
one million data points. 

Before describing the thermodynamic quantities
in BHB$\Lambda K^- \phi$  EoS table, we discuss the $\beta$-equilibrated matter
relevant to cold neutron stars. We generate the EoS of neutron stars by imposing charge
neutrality with the inclusion of electrons and the $\beta$-equilibrium
condition without neutrinos at temperature T = 0. In Fig.1a, various particle fractions are plotted as a function of baryon number density. The NS core contains a high neutron fraction. The proton fraction increases monotonically as baryon
density increases. The positive charges of protons are balanced
by negative charges of electrons. $\Lambda$-hyperons appear at 
$\sim 2.2 n_B$ followed by the $K^-$ condensate. The early appearance of $\Lambda$ hyperons makes the EoS softer delaying the appearance of the antikaon condensate.
The $K^-$ fraction in the condensate abundance reaches a considerable fraction at $3.5 n_B$. As soon as $K^-$ appears, the $e^-$ fraction drops and the charge neutrality is totally taken care off by the 
negatively charged condensate at higher density. On the other hand the population of neutrons is arrested due to the accumulation of more $\Lambda$s at higher density.

The mass–radius relationship of the sequence of neutron stars
is shown in Fig.1b for  HS(DD2), HS(DD2)$K^-$, BHB$\Lambda\phi$ and BHB$\Lambda K^- \phi$  EoS.
The solid  line (green) represents the nucleons-only neutron star. On the other hand,  dotted (magenta), dashed-dotted (blue) and 
dashed (red) lines represent neutron stars with additional $K^{-}$ condensate only, $\Lambda$-hyperons only, and both $\Lambda$-hyperons and
$K^{-}$ condensate respectively. The maximum masses for the four EoSs are 2.42, 2.24, 2.1 and 2.05 $M_{solar}$,  their corresponding radii being 11.89, 12.0, 11.58, 11.62 km respectively. The $\Lambda$- hyperons make
the EoS softer, resulting in a smaller as well as lighter maximum mass neutron
star compared with that of the nucleons-only
HS(DD2) EoS. The maximum mass is further
lowered in the presence of $K^-$ condensate.
It is to be noted that the masses are well within the observational benchmark of measured 2 M$_{solar}$ neutron stars \cite{anto,croma}.

We now move on to the results relevant to the 
BHB$\Lambda K^- \phi$ 
EoS table for supernova and NS merger simulations.
%In the present work, all Fermi–Dirac integrals
%are solved with the very accurate and efficient methods of
%Aparicio (1998) and Gong et al. (2001), complemented by
%analytic approximations where these are even more reliable, %whereas the Bose-Einstein integrals are computed by %modified Gaussian quadrature method where the limits  $0$ %to $\infty$ are mapped to $0$ to $1$.
Fig. \ref{fig:1} exhibits the composition of supernova matter as a function of baryon
mass density for T = 1, 50, and 100 MeV and $Y_q$ = 0.1, 0.3,
and 0.5. Number densities of various particles, such as  light(Z $\leq 5$),  heavy nuclei ($Z \geq 6$), neutrons, protons, $\Lambda$-hyperon and antikaons, both thermal and condensate are plotted. The heavy nuclei exist only at the low temperature and low density non-uniform matter.  The light nuclei on the other hand appear in the low density region only for T=50 MeV and higher charge fractions, as evident from the middle panel of Fig. \ref{fig:1}. At very high temperatures $> 48$ MeV, nuclei do not appear. Furthermore, nuclei dissolve into their fundamental constituents around normal nuclear matter density and form a uniform matter of neutrons and protons. The $\Lambda$-hyperons  appear with significant abundance($>10^{-5}$) at higher density at the cost of neutrons. Higher the temperature, lower is their  threshold density. However the population fractions of baryons do not differ with temperature at relatively higher density. Next we focus on the (anti)kaon number density. The antikaon condensate does not appear at all for a system with higher charge fraction $Y_q$. Even for $Y_q=0.1$, its threshold is shifted towards higher density, and it fails to appear at T=100 MeV. This may be attributed to the abundance of thermal kaons at 
higher temperatures, which cannot be traced at low temperature regime.
In fact above the critical temperature, the  condensate disappears producing thermal (anti)kaons.  Nevertheless, $\Lambda$-hyperons also play a dominant role. We have reported in an earlier work with (anti)kaons, but no hyperons, the density of
$K^-$ mesons in the condensate even dominates over that of thermal (anti)kaons ($n^{T}_{K}$) at T=50 MeV for $Y_q =0.1$ and $0.3$ \cite{EPJC}. Here this trend
is noted only for $Y_q =0.1$. Thus we conclude that the $\Lambda$-hyperons delay or do not allow the $K^-$ condensate to appear in the 
system, more so at higher $Y_q$.
\begin{figure}
\centering
\begin{minipage}{.45\textwidth}
    \centering
    \includegraphics[width=0.8\columnwidth, height=8cm]{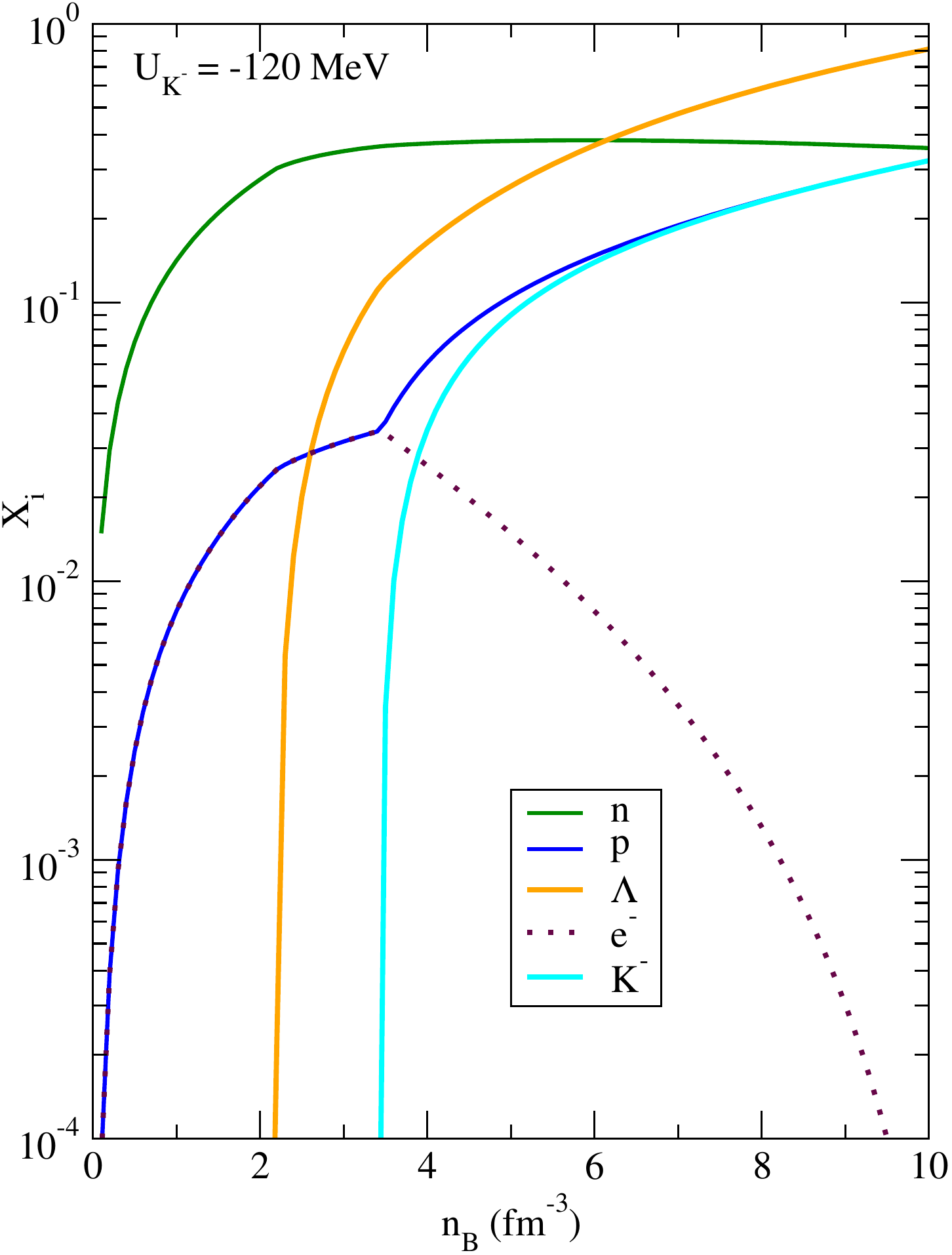}
%    \caption{Particle fractions of various species in a $\beta$-equilibrated cold NS are plotted as a function of baryon number  density for BHB$\Lambda K^- \phi$  EoS}
    \label{fig:frac}
    \end{minipage}
    \begin{minipage}{.45\textwidth}
    \centering
    \includegraphics[width=0.8\columnwidth, height=8cm]{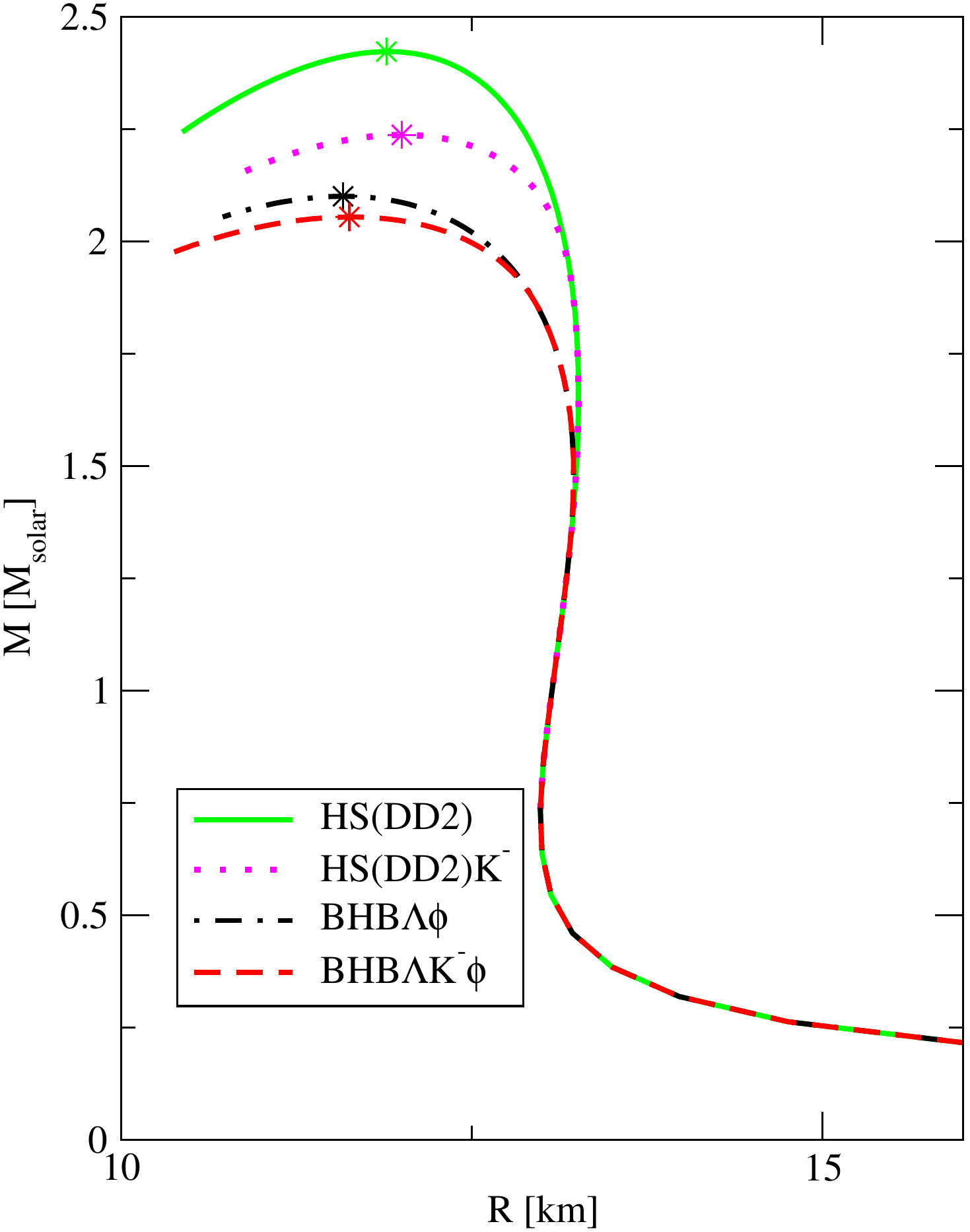}
%    \caption{Masses of the NS sequence are plotted as a function of radius for nucleonic and strange EoSs. Stars mark the maximum mass configurations.}
    \label{fig:mr}
    \end{minipage}
    \caption{Properties of $\beta$-equilibrated cold neutron star: a)Particle fractions of various species in a $\beta$-equilibrated cold NS are plotted as a function of baryon number  density for BHB$\Lambda K^- \phi$  EoS b)Masses of the NS sequence are plotted as a function of radius for nucleonic and strange EoSs. Stars mark the maximum mass configurations.}
\end{figure}

\begin{figure}
    \centering
    \includegraphics[width=0.8\columnwidth, height=9cm]{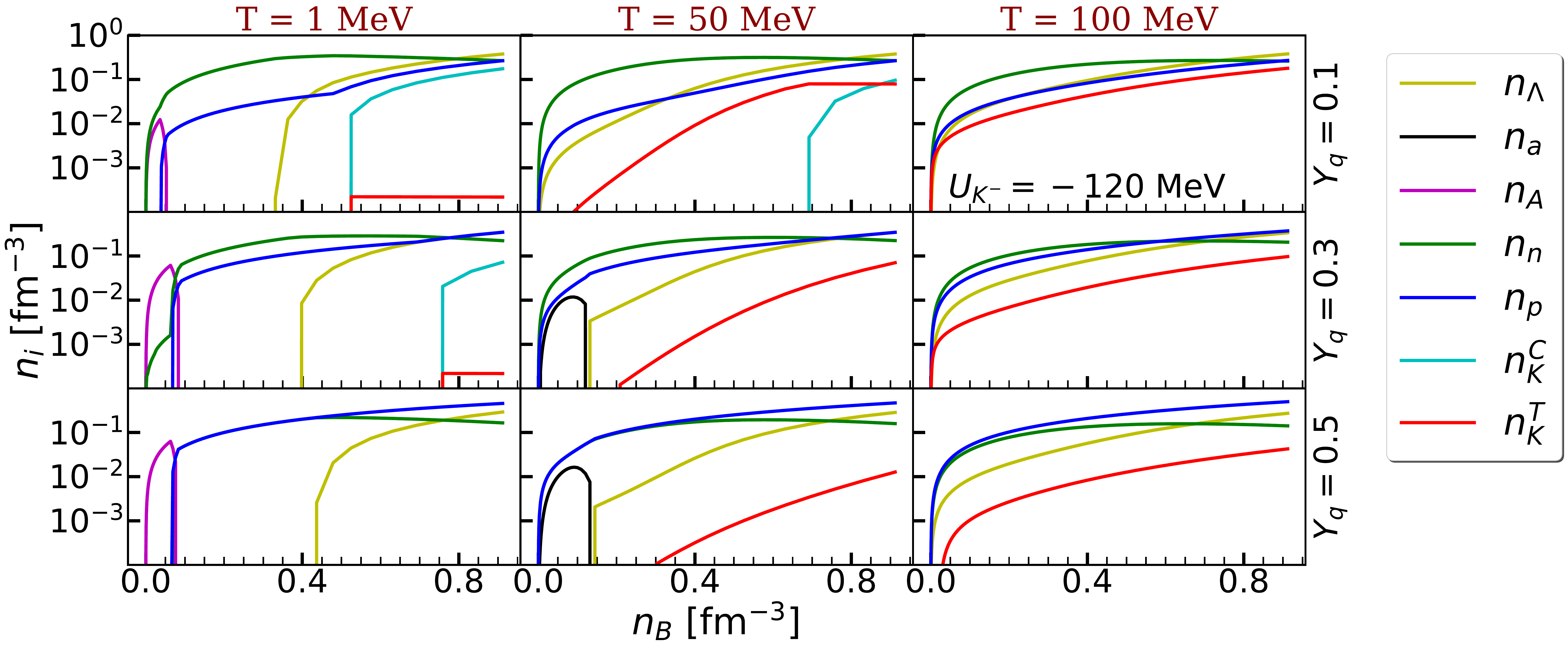}
    \caption{Number densities of different particle species, such
as light \& heavy nuclei, neutrons, protons and antikaons, both thermal and
condensate, $\Lambda$-hyperons as a function of baryon
number density for T=1, 50, 100 MeV and $Y_q$ = 0.1, 0.3, 0.5.}
    \label{fig:1}
\end{figure}

\begin{figure}
    \centering
    \includegraphics[width=1.0\columnwidth]{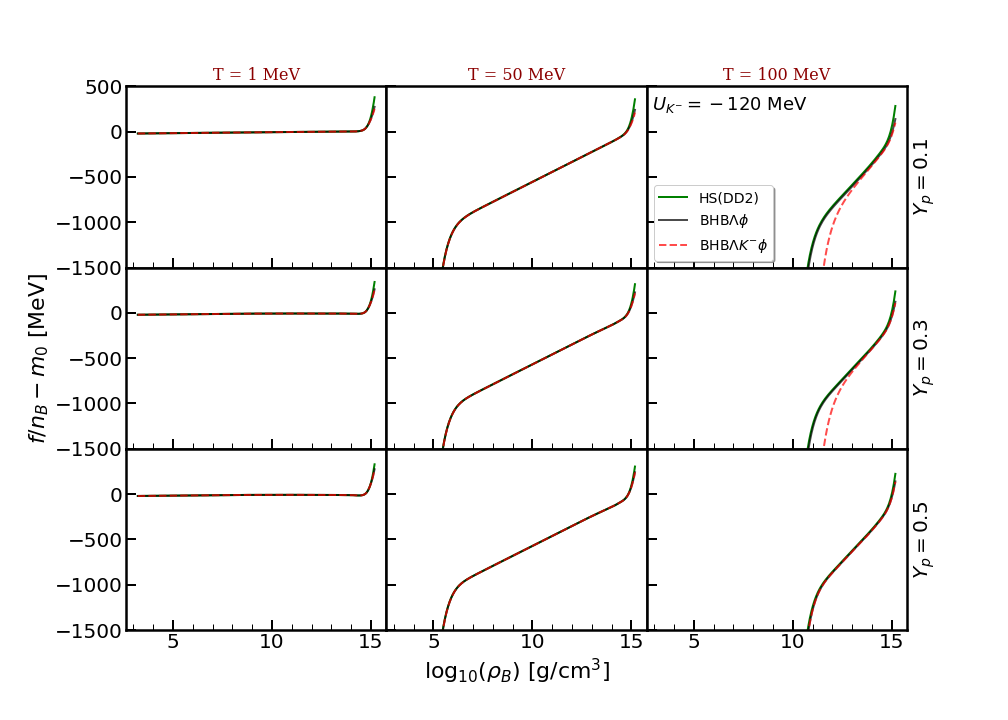}
    \caption{Free energy per baryon with respect to $m_0 = 938$ MeV is plotted as a function of baryon mass density for temperatures T = 1, 50, 100 MeV and positive charge
fraction $Y_q$ = 0.1, 0.3 and 0.5. Results from the nucleonic EoS table HS(DD2) (green) and strange EoS tables BHB$\Lambda\phi$ (black)  BHB$\Lambda K^- \phi$  (red-dashed) are
shown here}
    \label{fig:2}
\end{figure}

We have studied various thermodynamic observables
such as free energy per baryon, entropy per baryon and pressure as a function of baryon mass density. In all following figures, we
only show the hadronic contribution and plot them for various regimes of temperatures, T = 1,
50, and 100 MeV and positive charge fractions, $Y_q$ = 0.1, 0.3, and 0.5. Free energy per baryon with respect to the arbitrary value of $m_0$ = 938 MeV is plotted for BHB$\Lambda K^{-} \phi$  EoS in Fig. \ref{fig:2} as a function of mass density(in red dashed line).
The results for nucleons-only HS(DD2) EoS(green solid) and BHB$\Lambda \phi$ (black solid line)EoSs are also drawn for comparison. At lower
densities, there is practically no difference between the results of
nuclear and strange matter for different situations considered. A slight difference is observed at higher density for T=1 and 50 MeV, when the strange particles appear. The difference is quite prominent at T=100 MeV, 
%when a significant fraction of strange particles are present
owing to appearance of thermal (anti)kaons at a lower density along with the $\Lambda$s. 
If we  compare the the right-most panel of Fig. \ref{fig:2} at T=100 MeV, the effect of thermal (anti)kaons become evident. The difference between free energy of nuclear and strange matter is noticed to disappear apparently for $Y_q$=0.5. This may be attributed to the dissolution of the condensate with the production of thermal (anti)kaons at comparatively higher density. Also its abundance is observed to be relatively small compared to the matter with lower $Y_q$.

\begin{figure}
    \centering
    \includegraphics[width=1.0\columnwidth, height=10cm]{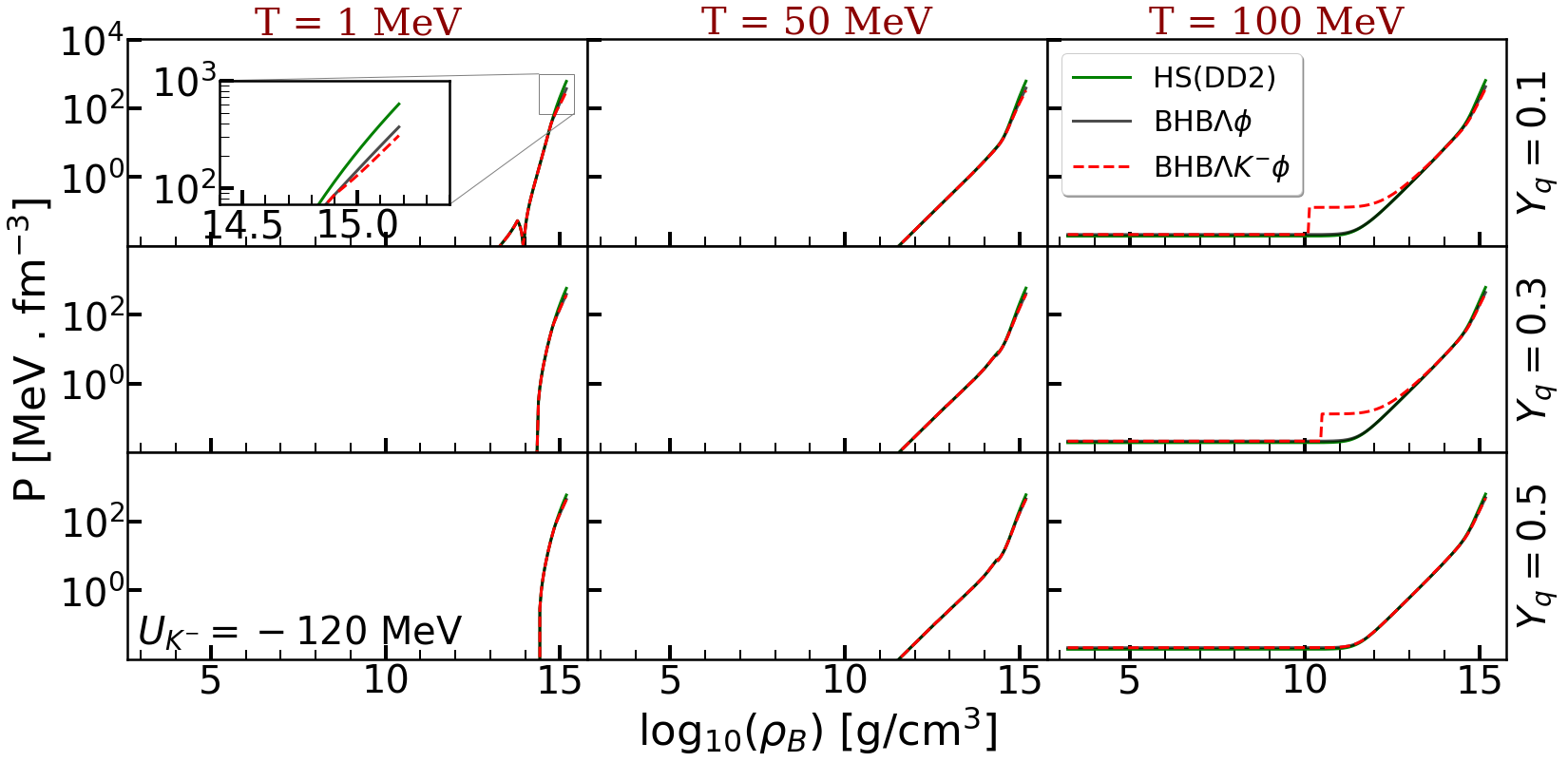}
    \caption{Pressure is shown as a function of baryon mass density for T=1, 50,
100 MeV and $Y_q$ = 0.1, 0.3, 0.5. Results are shown for three EoSs as depicted in \ref{fig:2}.The difference in the EoS at higher density is exhibited in the inset for $Y_q$ = 0.1 and T=1 MeV.}
    \label{fig:3}
\end{figure}

 The pressure is plotted in Fig. \ref{fig:3}, as a function of
baryon mass density. Just like the free energy case,
we find the nuclear and strange EoSs  do not show any difference at low densities for  the different values of temperature and positive charge fractions considered here. However, at the higher density the exotic particles appear, which clearly makes
the BHB$\Lambda K^- \phi$  EoS softer compared to the HS(DD2) and  BHB$\Lambda\phi$ EoSs.
The high density portion is zoomed in the inset box for
$Y_q=0.1$ and T=1 MeV to highlight this difference. A kink in pressure is observed at $\sim 10^{10}$ $g/cm^3$
for T=100 MeV and $Y_{q} =0.1$ and 0.3, which is due to significant contribution of thermal $K^{-}$ mesons to the pressure. There is no kink or jump in pressure when $\Lambda$s appear in the
system indicating  a smooth transition from nuclear
to hyperon matter.

\begin{figure}
    \centering
    \includegraphics[width=1.0\columnwidth, height=10cm]{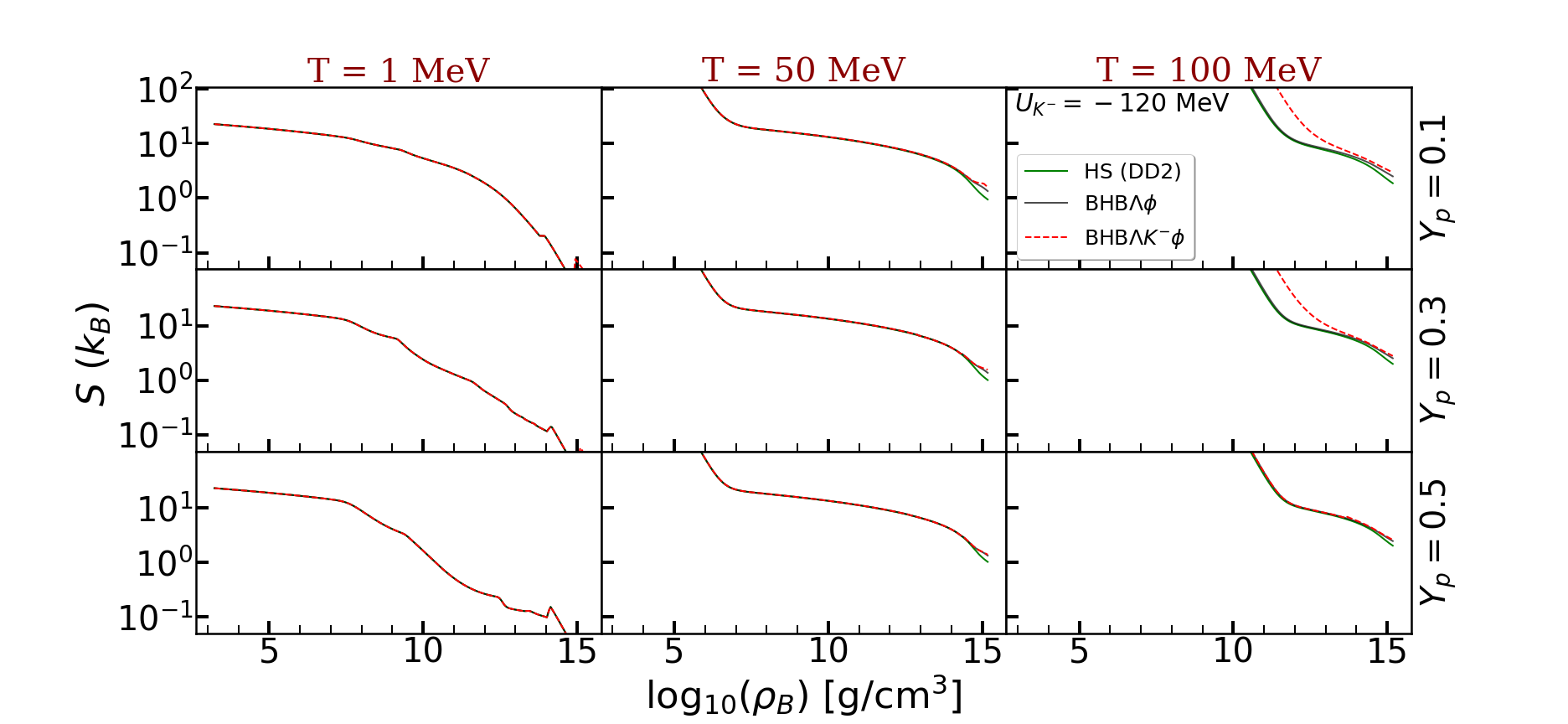}
    \caption{Entropy per baryon is shown as a function of baryon mass density for temperatures T = 1, 50, 100 MeV and positive charge fractions $Y_q$ = 0.1, 0.3, and 0.5. Results
from the nucleonic EoS table HS(DD2) (green) and two strange matter EoS tables BHB$\Lambda\phi$ (black ) and 
BHB$\Lambda K^- \phi$ (red dashed) are shown here.}
    \label{fig:4}
\end{figure}

In Fig. \ref{fig:4} the 
entropy per baryon
as a function of baryon mass density for the same set of 
values of temperatures and positive charge fractions are plotted. Here also, the effect of strange particles become evident in the high density region.
The kinks at low densities and temperatures
originate from changes in the nuclear composition which are
related to nuclear shell effects, whereas the kinks at higher densities mark the appearance of exotic particles. The difference between the three EoSs is prominent only at higher density for T=50 MeV. However the large fraction of  thermal (anti)kaons at low density does make a significant 
effect at T=100 MeV for $Y_q$= 0.1 and 0.3. This effect is blurred at $Y_q$=0.5, as the thermal (anti)kaons appear at higher density and the BHB$\Lambda K^- \phi$ curve deviates from HS(DD2) and BHB$\Lambda \phi$ EoSs at the higher density end.

\begin{figure}
    \centering
    \includegraphics[width=0.6\columnwidth, height=9cm]{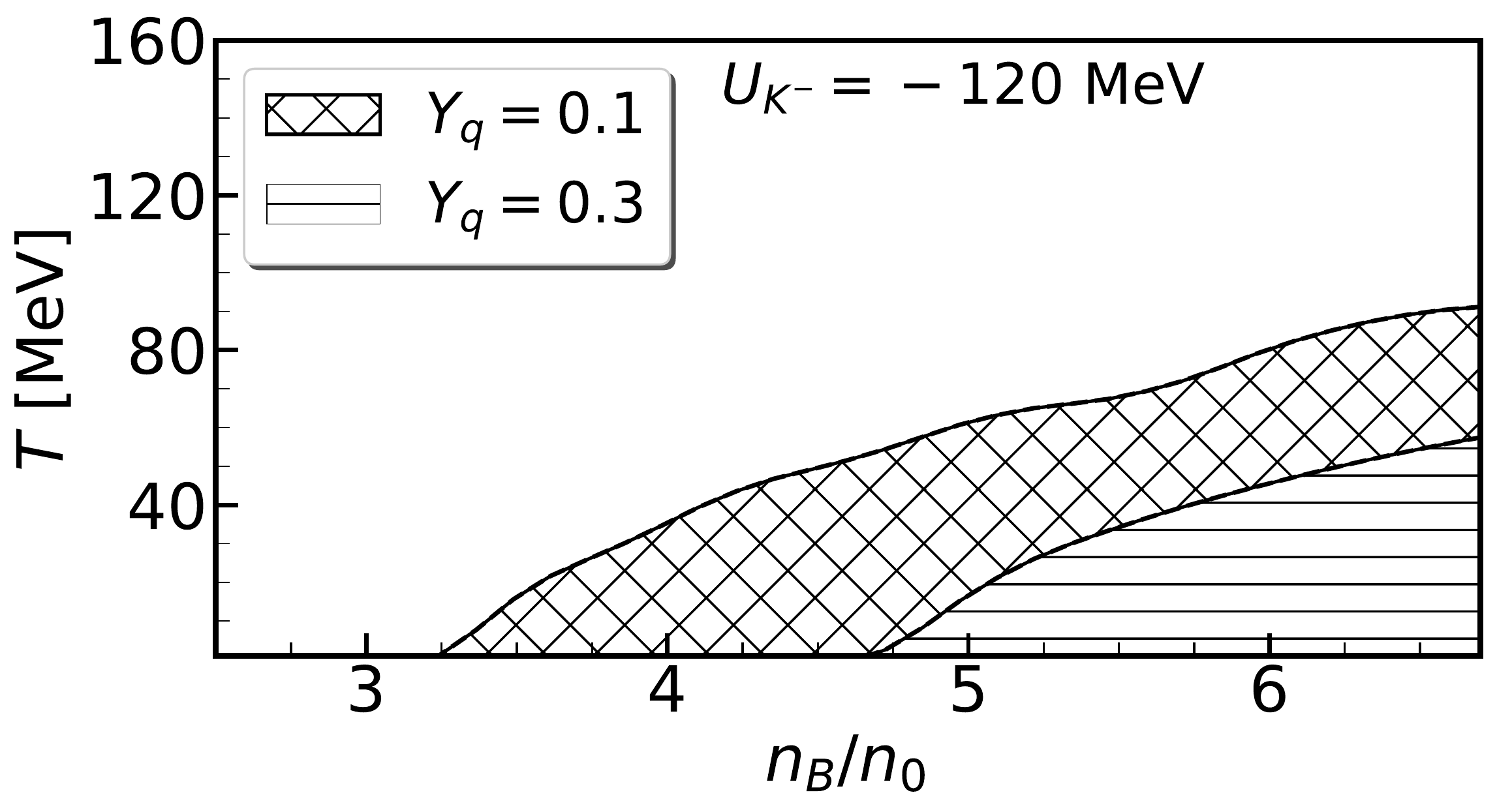}
    \caption{The phase diagram of hyperon matter with antikaon condensate is
displayed for $Y_q$ = 0.1, 0.3, and U$_{K{^-}}= -120$ MeV.
The $K^{-}$ condensed phases, represented by the shaded regions are demarcated
from the hadron phases by the solid lines for the two $Y_q$ values.}
    \label{fig:5}
\end{figure}

Finally in Fig. \ref{fig:5}, we plot the phase diagram, i.e. the temperature versus 
baryon density for $Y_q$=0.1 and 0.3.
The shaded regions below the solid lines represent the kaon condensed phase in 
the presence of hyperons. 
The solid line demarcates the hyperon  matter phase from the condensate phase 
for a particular value of $Y_q$ and denotes the critical temperature of kaon 
condensation at each density point. 
At higher $Y_q$, the condensate appears at higher density. Condensate may not appear at all if the temperature of the system is high, as was seen in 
Fig. \ref{fig:1}. 
\cite{EPJC} exhibited a similar phase 
diagram of nuclear matter with antikaon condensate. Comparing this phase 
diagram with 
that of \cite{EPJC}, clearly shows the effect of hyperons. The hyperons
visibly delays the kaon condensation to a higher density. Also the critical 
temperature, above which
the $K^-$ condensate dissolves, comes down at a given density. The condensed 
phase is observed to shrink with higher $Y_q$, in nuclear as well as hyperon 
matter; 
it ceases to exist at all in the presence of hyperons
at $Y_q=0.5$. 

\section{Summary \& conclusion} \label{sec:floats}
We have constructed a new EoS table including (anti)kaons and hyperons for core collapse supernova and binary NS 
merger simulations. We call it HS(DD2)$\Lambda K^- \phi$. The low density, non-uniform matter  of this EoS table is
generated within an extended version of the NSE model with
excluded volume \cite{hs1}.  The uniform
matter on the other hand is described in a finite temperature density-dependent relativistic mean field model. The
nucleons are described within widely-used meson exchange model using the DD2 parameter set  \cite{typ10}.
The interaction between $\Lambda$-hyperons is mediated by
additional $\phi$ mesons. It is noted that
the fraction of $\Lambda$ hyperons is significant at $\sim 2$ times normal nuclear matter density for a cold matter. At higher temperature $\Lambda$s populate even at low density matter, their population  grows in
uniform matter at the cost of neutrons at high density, eventually overshooting neutron fraction at very large density.

We have also considered the role 
of thermal (anti)kaons and the $K^-$ condensate on the EoS and other thermodynamic observables. At low temperature and low positive charge fraction, the
system is populated with $K^-$ condensate and a very small amount of thermal (anti)kaons. However, a high fraction of $\Lambda$s at low density does not favour the onset of $K^-$ condensate. At high temperature, only the thermal (anti)kaons populate the matter. 

The presence of  $\Lambda$ hyperons, thermal (anti)kaons  and the
antikaon condensate makes the HS(DD2)$\Lambda K^- \phi$ EoS softest of the nucleon-only
HS(DD2) EoS,  and the strange matter EoS HS(DD2)$\Lambda \phi$, and  the HS(DD2)K$^-$ EoS.

We have shown various thermodynamic observables
such as the free energy per baryon, entropy per baryon and pressure of
HS(DD2)$\Lambda K^- \phi$ matter as a function of baryon mass density for the set of  temperatures, T = 1,
50, and 100 MeV and positive charge fractions, $Y_q$ = 0.1, 0.3, and 0.5. For comparison we also plot
the corresponding quantities of HS(DD2) and 
HS(DD2) $\Lambda \phi$ matter. The effect of the strange particles is mainly observed at high density, that becomes visibly prominent at high temperature. 

Finally, we also 
compute the EoS of the charge neutral and $\beta$-equilibrated
cold EoS and report the lowering of maximum mass neutron star in the presence of strange particles. The maximum masses are  2.42, 2.24, 2.1 and 2.05 $M_{solar}$, all are compatible with the heavy neutron stars of masses $\sim 2 M_{solar}$,
discovered in the past decade \cite{anto,croma}. Also, the radii match with the range of radius calculated from the tidal deformability 
%($70 \leq {\Lambda_{1.4}} \leq 580$) 
extracted from the BNS merger GW170817 event for a 1.4$M_{solar}$ neutron star\cite{ligo1}.  
Recently the Neutron Star Interior Composition Explorer (NICER) observation of PSRJ0030+0451 has come up with a radius estimation  of $13.02^{+1.24}_{-1.06}$ km for the 1.44$M_{solar}$ pulsar PSR J0030+0451 \cite{miller, watts}. Our results for radius corresponding to 1.4M$_{solar}$ neutron star is slightly higher than that of GW170817 \cite{soma} whereas it is consistent with the NICER
observation.

We shall perform supernova simulations and neutron star merger simulations with new HS(DD2)$\Lambda K^- \phi$ 
EoS table and leave them for a future publication.

\newpage
\appendix
\section{Description of the table}
We follow the format of the widely used, existing supernova EoS tables -Shen \cite{Shen} and BHB$\Lambda \phi$ \cite{bhb}. We arrange the data
in a parameter grid of temperature, density and positive charge fraction. The first two parameters have a logarithmic spacing, while the charge fraction is on a linear scale. We group them in blocks of
a fixed temperature, starting with the lowest value. Within each
temperature block, we group the data according to the positive charge fraction, again starting with lowest values. Finally, for a given temperature
and positive charge fraction, we list all the thermodynamic properties according to ascending  baryon number densities.

 We record only hadronic contributions to 
different quantities in the table. The contributions of photons, electrons, 
positrons and neutrinos can be added separately. There are twenty-three entries of the table
representing different
thermodynamic quantities corresponding to each density grid point in the 
table. These thermodynamic quantities are listed below.

1.Logarithm of baryon mass density (log$_{10} (\rho_B)$ [g/cm$^3$]) 
is defined as the baryon number density multiplied by 
the value of the atomic mass unit $m_u = $ 931.49432 MeV.

2.Baryon number density ($n_B$ [fm$^{-3}$])

3.Logarithm of total positive charge fraction (log$_{10} (Y_p)$) 

4.Total positive charge fraction ($Y_p$)

5.Free energy per baryon ($F$) relative to 938 MeV is defined by
\begin{equation}
F = \frac{f}{n_B} - 938~.
\end{equation}

6.Internal energy per baryon ($E_{int}$) relative to $m_u$ is defined by
\begin{equation}
E_{int} = \frac{\epsilon}{n_B} - m_u~,
\end{equation}
where the energy density $\epsilon$ is given by Eq.({\ref{eqn:ed}}).

7.Entropy per baryon ($S$ [$k_B$])
related to the entropy density through $S = \frac{\mathcal S}{n_B}$.

8.Average mass number of heavy nuclei ($\bar A$) is defined as
$\bar A = \sum_{A,Z \geq 6}A n_{A,Z} / \sum_{A,Z\geq 6} n_{A,Z}$ 

9.Average charge number of heavy nuclei ($\bar Z$) is defined as
$\bar Z = \sum_{A,Z \geq 6}Z n_{A,Z} / \sum_{A,Z\geq 6} n_{A,Z}$ 

10. Nucleon effective mass ($m^*$ [MeV])

In the RMF calculation, we use separate values for neutron and proton masses, 939.56536 \&  938.27203 MeV respectively. 
However,  the average value of neutron and proton effective masses is stored.

%11.Effective mass of proton ($m_p^*$ [MeV])

%12.Effective mass of $\Lambda$ ($m_{\Lambda}^*$ [MeV])

11.Mass fraction of unbound neutrons ($X_n=n_n/n_B$) 
%This is defined as $X_n = $.

12.Mass fraction of unbound protons ($X_p=n_n/n_B$) 

13.Mass fraction of unbound $\Lambda$s ($X_{\Lambda}=n_{\Lambda}/n_B$)

14.Mass fraction of light nuclei ($X_a$) is defined as 
$X_a = \sum_{A,Z \leq 5}A n_{A,Z} / n_B$ 

15.Mass fraction of heavy nuclei ($X_A$) is defined as 
$X_A = \sum_{A,Z \geq 6}A n_{A,Z} / n_B$ 

16.Baryon pressure ($P$ [MeV/fm$^3$])

17.Neutron chemical potential relative to neutron rest mass ($\mu_n-m_n$ [MeV]). Whenever $\Lambda$'s are present in the system,  $\mu_n = \mu_{\Lambda}$ condition is satisfied.

18.Proton chemical potential relative to {proton} rest mass ($\mu_p-m_p$ [MeV])

%19.$\Lambda$ chemical potential relative to {$\Lambda$} rest mass ($\mu_{\Lambda}-m_{\Lambda}$ [MeV])

19.Average mass number of light nuclei ($\bar a$) is defined as
$\bar a = \sum_{A,Z \leq 5}A n_{A,Z} / \sum_{A,Z\leq 5} n_{A,Z}$ 

20.Average charge number of light nuclei ($\bar z$) is defined as
$\bar z = \sum_{A,Z \leq 5}Z n_{A,Z} / \sum_{A,Z\leq 5} n_{A,Z}$ 

21.Kaon effective mass  ($m^*_K$ [MeV])

22.Mass fraction of thermal kaon ($X_K^T=n_K^T/n_B$)  

23.Mass fraction of $K^-$ condensate ($X_K^C=n_K^C/n_B$)

%% ************** eDIT Tuhin @ 16 Jan 2021

%\section{Appendix information}

\leftline {\bf Acknowledgements}
Authors acknowledge the DAE-BRNS grant received under the BRNS project No.37(3)/14/12/2018-BRNS. DB acknowledges the hospitality at FIAS and support of Alexander von Humboldt Foundation, Germany. The calculations are performed in the server of Physics Department, BITS-Pilani, Hyderabad Campus.\\

\bibliographystyle{aasjournal}



\end{document}